# Secure CDMA Sequences

Anatolii Leukhin, Oscar Moreno and Andrew Tirkel

*Podvolgskii State Technical University, Gauss Research Foundation, Scientific Technology*

++7 (8362) 68-78-42, +1 (787) 689-5869, +61-395922206

+7 (8362) 45-53-73, +1 (787) 689-5867, +61-395922206

leukhinan@list.ru, moreno@nic.pr, atirkel@bigpond.net.au

Abstract: Single sequences like Legendre have high linear complexity. Known CDMA families of sequences all have low complexities. We present a new method of constructing CDMA sequence sets with the complexity of the Legendre from new frequency hop patterns, and compare them with known sequences. These are the first families whose normalized linear complexities do not asymptote to 0, verified for lengths up to $6\times 10^8$. The new constructions in array format are also useful in watermarking images. We present a conjecture regarding the recursion polynomials.

We also have a method to reverse the process, and from small Kasami/No-Kumar sequences we obtain a new family of $2^n$ doubly periodic $(2^n+1)\times(2^n-1)$ frequency hop patterns with correlation 2.

*CDMA, correlation, linear complexity, array, shift sequence, frequency hop*

## Introduction

Mobile communications require flexible networks which can accommodate variable numbers of users under adverse propagation environments. Multiplexing by frequency or time division is inefficient and ineffective in guarding against multipath and mutual interference, noise and jamming. Code division multiple access (CDMA), frequency hopping, orthogonal frequency division multiplexing (OFDM), and ultra wide band (UWB) techniques have been developed to overcome these shortcomings. These have been enhanced by multiple input, multiple output (MIMO), polarization and space-time coding. All rely on digital sequences with good correlation. Ensembles of sequences with low off-peak autocorrelation and low cross-correlation are used in asynchronous CDMA systems. Such sequences are statistically uncorrelated, and the sum of a large number of sequences results in multiple access interference (MAI) that is approximates Gaussian noise, due to the central limit theorem. An important feature of such multiple-access codes is their cryptographic security. A standard measure of security is the linear complexity, as computed using the Berlekamp-Massey algorithm, or its generalization, the Reeds-Sloane algorithm. A more useful measure is the complexity normalized to the sequence length, i.e.

normalized complexity. Sequence sets commonly used in CDMA include Gold, Kasami, Kerdock and Bent. In general, their linear complexity is low, and the normalized complexity asymptotes to 0 as the sequence length increases. Among the known constructions, the No-Kumar sequences, a generalization of the small Kasami sequences, have the highest linear complexity. In this paper, we present a new method of modifying the linear complexity of the No-Kumar sequences. Also, we analyze the new Moreno-Tirkel sequence sets, whose normalized complexity asymptotes to 0.5 or 1.0. The small Kasami, No-Kumar and Moreno-Tirkel sequence sets can also be represented as sets of doubly periodic arrays, where the columns are cyclic shifts of a pseudonoise sequence or constant columns. The small Kasami and No-Kumar sequences inherit this property from their parent long m-sequence, and their columns are cyclic shifts of a short m-sequence. The Moreno-Tirkel sequences are constructed as arrays, and then unfolded from arrays into sequences using the Chinese Remainder Theorem. The array interpretation is useful in analyzing the correlation and complexity of these sequence sets. It also makes it easy to generalize the constructions, by substituting column sequences with greater complexity. All the arrays in this paper can also be used in digital watermarking of images. In addition, it is possible to transform these arrays into frequency hopping patterns with good auto and cross-correlation, The paper is organized as follows. Section 2 presents the definitions and necessary preliminaries, including the array interpretation and a review of the Moreno-Tirkel construction. Section 3 presents linear complexity results for all the sequence sets discussed above for the lengths $2^8-1$ and $2^{12}-1$. Conclusions are presented in Section 4.

## PRELIMINARIES

Periodic shift by $\tau$ for a sequence $s$. The sequence $s^\tau$, the shifted sequence of $s = [s_i], i = 0,1,,..., L-1$ by $\tau$ places to the left, is defined as $s^\tau = [s_{i+\tau}], i = 0,1,,..., L-1$, ($i+\tau$ is calculated modulo $L$). The $n^{th}$ decimation of a sequence $s = [s_i], i = 0,1,,..., L-1$ is given by $dec_n(s) = s_{ni \,(\bmod L)}$ for $i = 1,2,..., L-1$. The periodic shift $(h,v)$ of an array A. For any array $A = [a_{ij}], 0 \le i \le m-1, 0 \le j \le n-1$, the shifted array of A by $h$ places horizontally to the left, and $v$ places vertically downwards, denoted by $A^{(h,v)}$, has (i,j) entry equal to $a_{i+h,j+v}$, where $i+h$ and $j+v$ are calculated modulo $m$ and $n$ respectively. The periodic autocorrelation, $\theta_s(\tau)$, of a sequence $s$ for shift $\tau$ is given by

$$\theta_s(\tau) = \sum_{i=0}^{L-1} s_i s_{i+\tau}^*, \qquad (1)$$

where $L$ is the length of $s$, $i+\tau$ is calculated modulo $L$. In terms of the dot product, the autocorrelation value $\theta_s(\tau) = s \cdot s^\tau$. The periodic autocorrelation, $\theta_A(h,v)$, of the $m \times n$ array, $A = [a_{ij}]$, for horizontal shift $h$ and vertical shift $v$ is given by

$$\theta_A(h,v) = \sum_{i=0}^{m-1} \sum_{j=0}^{n-1} a_{ij} a_{i+h,j+v}^* \qquad (2)$$

where $i+h$ and $j+v$ are calculated modulo $m$ and $n$ respectively.

For $h=v=0$ the autocorrelation value $\theta_A(0,0)$ is called the peak value, and for $(h,v) \neq (0,0)$, the values $\theta_A(h,v)$ are called off-peak values. The periodic cross-correlation, $\theta_{AB}(h,v)$, of the $m \times n$ arrays, and $B = [b_{ij}]$, for horizontal shift $h$ and vertical shift $v$ is given by

$$\theta_{AB}(h,v) = \sum_{i=0}^{m-1} \sum_{j=0}^{n-1} a_{ij} b_{i+h,j+v}^* \qquad (3)$$

where $i+h$ and $j+v$ are calculated modulo $m$ and $n$ respectively.

**Linear Complexity**

Let $s^n = s_0 s_1 \ldots s_{n-1}$ be a sequence over a field $F$. The linear complexity or linear span of $s^n$ is defined to be the shortest positive integer $l$ such that there are constants $c_0 = 1, c_1, \ldots, c_l \in F$ satisfying

$-s_i = c_1 s_{i-1} + c_2 s_{i-2} + \ldots + c_l s_{i-l}$, for all $0 \leq i < n$.

Such a polynomial $c(x) = c_0 + c_1 x + \ldots + c_l x^l$ is called the *feedback polynomial* of a shortest linear feedback shift register (LFSR) that generates $s^n$. Hereafter we use feedback polynomial for short. It is not difficult to construct a solitary sequence of arbitrary linear complexity, by applying non-linear logical operations on the feedback coefficients of an LFSR. However, constructing a family of sequences with that property is more difficult. A shift register with a given feedback can produce different sequences with different cycle lengths, depending on the initial conditions. Here, we consider sequences of the same cycle length only. Also, there exist modifications to the Berlekamp-Massey algorithm which make it easier to decode sparse recursion polynomials. Hence, given two LFSR's of the same length, our preference is always for the one with greater Hamming weight. The exception is when the LFSR length is the same as the length of the sequence. For most sequences the linear complexity varies as the length increases. Therefore, it is useful to introduce the concept of normalized complexity i.e. $l/L$. The asymptotic normalized complexity is the limiting value as $L$ tends to infinity.

**Array interpretation**

*m-sequence, small Kasami*

We follow the exposition in [1]. One method of folding a pseudorandom sequence (PRS) into a two or higher dimensional array, sometimes referred to as a pseudorandom array (PRA), requires that the sequence length L be factorable into two or more relatively prime factors. If *L=uv*, with *gcd(u,v) =1*, then an array with u rows and v columns can be constructed by plotting the sequence down the main diagonal. When an edge is reached the array is reentered at the opposite figure on the next row or column.

| 1  | 29 | 57 | 22 | 50 | 15 | 43 | 8  | 36 |
|----|----|----|----|----|----|----|----|----|
| 37 | 2  | 30 | 58 | 23 | 51 | 16 | 44 | 9  |
| 10 | 38 | 3  | 31 | 59 | 24 | 52 | 17 | 45 |
| 46 | 11 | 39 | 4  | 32 | 60 | 25 | 53 | 18 |
| 19 | 47 | 12 | 40 | 5  | 33 | 61 | 26 | 54 |
| 55 | 20 | 48 | 13 | 41 | 6  | 34 | 62 | 27 |
| 28 | 56 | 21 | 49 | 14 | 42 | 7  | 35 | 63 |

(a) Filling Order

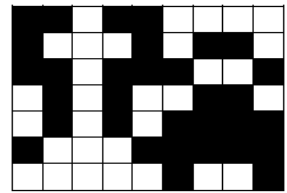

(b) 7×9 m-array

Fig. 1. Array Interpretation example

This ensures that the array is filled by a single pass through the sequence. The resulting array retains the autocorrelation structure of the original sequence when it is compared with any cyclic shift of itself in any combination of dimensions. The rows and columns of the array exhibit sequences which correspond to those produced by sampling the original sequence at a rate determined by u and v. In the special case when $L = 2^{2m} - 1 = (2^m + 1)(2^m - 1)$, a square array with $2^m - 1$ rows and $2^m + 1$ columns can be formed. The sampling values can be shown to be

$$R = 2^{m-1}(2^m - 1), \quad C = 2^{m-1}(2^m + 1) \qquad (4)$$

A simple method of obtaining 8 small Kasami arrays from the m-array in Fig 1(b) is by adding the columns of the m-array to the 7 different cyclic shifts of the m-sequence column, and appending the parent m-sequence. Note that the four cyclic shifts appearing in the m-array appear twice, or not at all, and the sequence of shifts is palindromic. Following the addition of an m-sequence column in a shift which does not exist in the parent, there are 4 arrays with no constant columns. The addition of an m-sequence column in a shift which exists in the m-array results in an array with two constant columns. There are 3 such cases. The original m-array completes the small Kasami set.

*No-Kumar*

A variation on the theme can be performed by applying a trace map to an intermediate field, followed by exponentiation and a final trace map to base field, just as in the GMW construction. This results in sequences with greater linear complexity, called No-Kumar sequences, which can also be plotted as arrays.

*Generalization*

The autocorrelation and cross-correlation functions of the small Kasami and the No-Kumar sequences take on values from the set $\{-1, (-2^{\frac{m}{2}}-1), (2^{\frac{m}{2}}+1)\}$. In array format, this corresponds to 1 column match, 0 columns match, or 2 columns match. A matching column contributes $2^{\frac{m}{2}}-1$ to the array correlation, whilst a mismatched column contributes -1. This remains unchanged if the m-sequence columns were replaced by any other pseudonoise columns of commensurate length, in the same cyclic shifts as the m-sequence columns in the original arrays. In this context, pseudonoise refers to two valued autocorrelation with the off-peak value being -1. Suitable columns are the binary Legendre and the Hall sequence.

*Moreno-Tirkel*

[2] presents a new method of synthesizing arrays and then unfolding them into sequence sets with similar correlation properties to the small Kasami and No-Kumar families. The Moreno-Tirkel sequence sets are based on algebraic constructions of families of shift sequences with correlation 2. Family A is based on an exponential quadratic shift sequence of length p-1 modulo p. Family B uses a rational function map over $F_{p^m}$ plus ∞ producing a shift sequence of length p+1 modulo p. Family C uses a reduced modulo method to produce a shift sequence of length $2^n+1$ modulo $2^n-1$. Examples of such shift sequences are shown in the

upper part of Fig. 2. They are respectively 7×6 and 7×8 and 7×9. The lower part of Fig. 2. shows the arrays following the substitution of the columns by cyclic shifts of m-sequences of length 7 as described in [2].

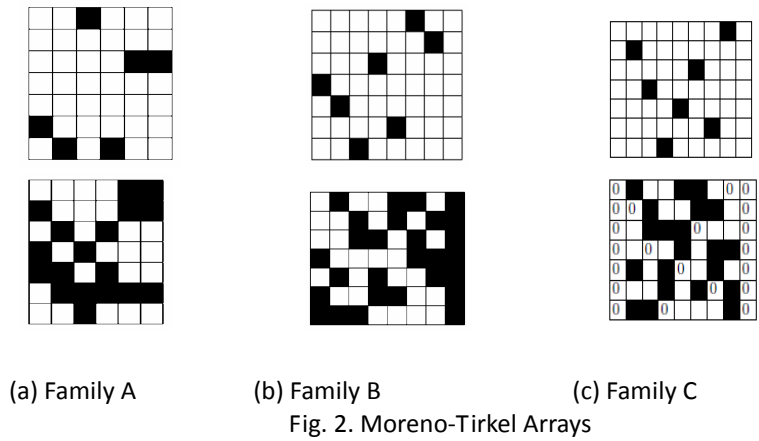

(a) Family A     (b) Family B     (c) Family C

Fig. 2. Moreno-Tirkel Arrays

**CDMA SEQUENCES IN ARRAY FORMAT**

Apart from the Kerdock sequences, the known CDMA sequences can be expressed in array format. Fig. 3. shows Small Kasami, No-Kumar, Gold and Bent sequences of length 255. The Small Kasami and No-Kumar exhibit symmetries, being cyclic shifts of an m-sequence column of length 15 and constant columns.

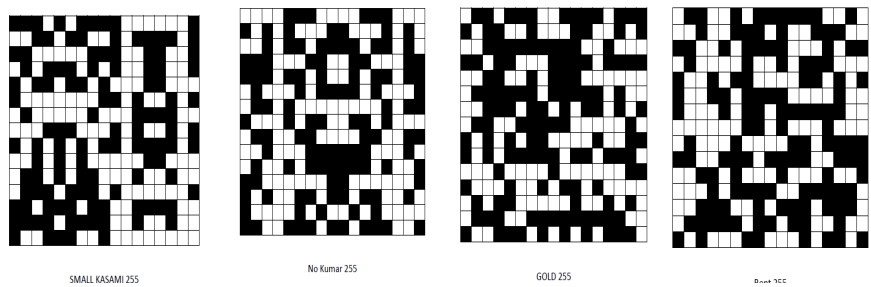

Fig. 3. Small Kasami, No-Kumar, Gold and Bent sequences of length 255 as 15×17 arrays

Larger arrays obtained by folding sequences of length 4095 into 63×65 format are shown in Fig. 4. They exhibit similar characteristics.

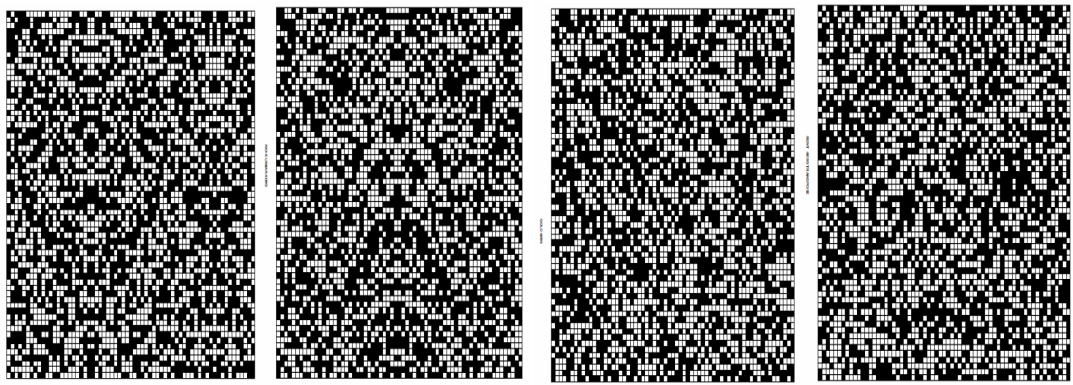

Fig 4. CDMA Sequences of length 4095 plotted as 63×65 arrays: Kasami, No-Kumar, Bent, Gold.

**GENERALIZATION**

Whenever the column length is commensurate with the m-sequence length, it is possible to replace those columns with identical shifts of any other pseudonoise sequence, with no change in the autocorrelation or cross-correlation. For No-Kumar sequences of length 1023 folded into 31×33 arrays, m-sequences of length 31 can be substituted by binary Legendre sequences of equal length. Fig. 5. shows the original array on the left, while the substituted array is on the right. It remains a mystery as to why the linear complexity following the substitution does not approach that of the column sequence, unlike the situation with the Moreno-Tirkel constructions discussed next.

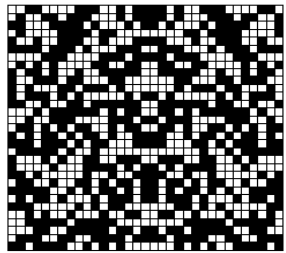  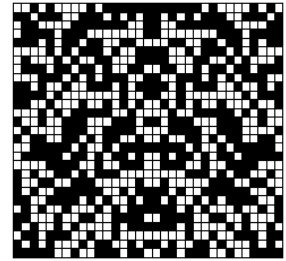

(a) Original with m-sequence columns        (b) Columns substituted by Legendre

Fig. 5. Generalization of No-Kumar sequences

**Moreno-Tirkel**

Fig. 6 shows the array form (18×19) of a sequence obtained from Family A, an array (19×20) from Family B, and a (15×17) array from Family C, where the rows have been substituted by binary Legendre sequences of length 17. These are the closest values to compare with the array format (15×17) of the previous known constructions. Note that there are no apparent symmetries.

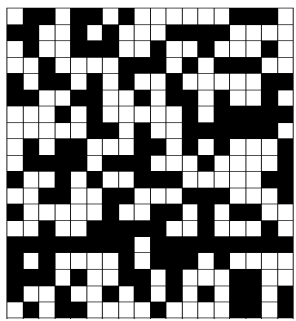  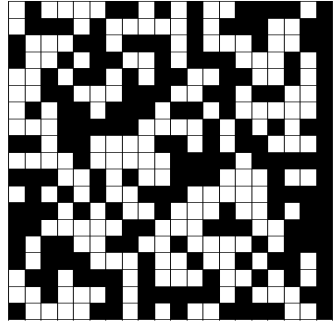  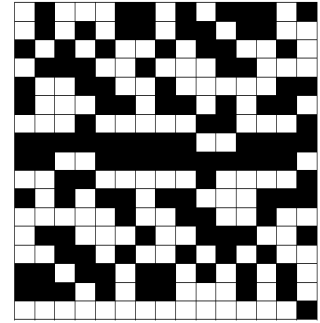

Family A 342        Family B 380        Family C 255

Fig. 6 Examples of Moreno-Tirkel arrays

The array of Family C is obtained by rotating the original array before substitution. This can be done because the original array has at most one dot per column and exactly one dot per row. The process is shown in Fig. 7.

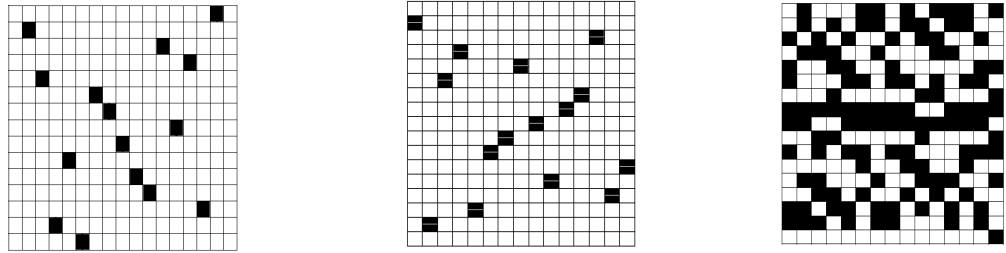

(a) Original Construction      (b) Rotated anticlockwise      (c) Substituted Array

Fig.7. Moreno-Tirkel Family C construction

**Frequency Hop Patterns**

The Moreno-Tirkel constructions use new frequency hop patterns to produce families of CDMA sequences. Now, we show how the converse applies to sequence families such as the small Kasami and No-Kumar. These are represented by arrays composed of cyclic shifts of a pseudonoise sequence or constant columns. The shift sequences which describe these arrays are doubly periodic, and can therefore be used as new families of frequency hop patterns. These families have auto and cross-correlation bound of 2. An example of this correspondence is shown in Fig. 8 for two Kasami sequences of length 63 as 7×9 arrays.

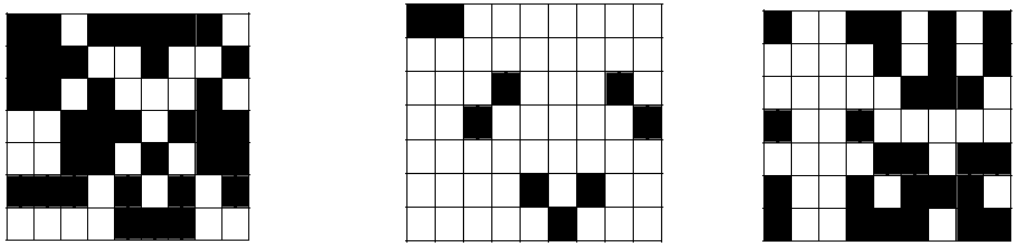

Frequency Hop Pattern A

Shift Sequence A: 0,0,3,2,5,6,5,2,3          Shift Sequence B: 5,-,-,5,6,4,0,4,6

0 shift is the start of the run of three black squares (arbitrary)

Fig. 8. Kasami Sequence conversion to Frequency Hop patterns

The rest of the family of frequency hop patterns is: C = 5,-,-,5,6,4,0,4,6. D = 5,-,4,-,5,0,2,2,0. E = 0,0,1,4,6,-,6,4,1. F = 0,2,3,5,3,2,0,1,1. G = 3,0,3,-,6,2,2,6,-. H = 5,3,3,5,2,6,4,6,2.

These new patterns can also be applied as time hopping patterns, optical orthogonal codes or sonar sequences.

# Linear Complexity Results

Linear complexity for known CDMA sequences and the three families of Moreno-Tirkel constructions with Legendre sequence columns are shown in Table 1.

| Family/Length | 255 | 1023 | 4095 | Asymptote |
|---|---|---|---|---|
| Bent [4] | 0.125 | - | 0.018 | $2^{-1.7(n-8)}$ |
| Small Kasami [5] | 0.047 | 0.015 | 0.004396 | $1.5n \times 2^{-n}$ |
| No-Kumar [6] | 0.235 | 0.152 | 0.031 | 0 |
| Generalized N-K | - | 0.132 | - | 0 |
| Gold [7] | 0.063 | 0.02 | 0.005861 | $n2^{-n+1}$ |
| Kerdock [8] | 0.142 | 0.054 | 0.019 | $2^{-1.4(n-8)}$ |
| Moreno-Tirkel Family A | 0.947 (length=342) | 0.485 (length=930) | 0.985 (length=4422) | $\approx 0.5$ for p=8k+3 or 8k+5<br>$\approx 1.0$ for p=8k±1 |
| Moreno-Tirkel Family B | 0.985 (length=380) | 0.46875 (length=992) | 0.970588 (length=4556) | $\approx 0.5$ for p=8k+3 or 8k+5<br>$\approx 1.0$ for p=8k±1 |
| Moreno-Tirkel Family C | 0.4706 (length=255) | | | $\approx 0.5$ for p=8k+3 or 8k+5<br>$\approx 1.0$ for p=8k±1 |

Table 1. Linear Complexity Comparison

*Conjecture (Moreno-Tirkel)*

For Family A and C of the Moreno-Tirkel construction, the recursion polynomial of the long sequence is obtained from that of the column sequence by raising each term to the power equal to the number of columns in the array. This has been verified for lengths up to $6 \times 10^8$.

*Example*: Consider the array of Fig. 7(c). The column is a binary Legendre sequence of length 17: 0,1,1,0,1,0,0,0,1,1,0,0,0,1,0,1,1.

Column recursion polynomial: $x^8 + x^7 + x^6 + x^4 + x^2 + x^1 + 1$ [9]. Recursion polynomial for long sequence: $x^{120} + x^{105} + x^{90} + x^{60} + x^{30} + x^{15} + 1$. The number of columns in the array is 15.

**Correlation, Length and Set Size**

All CDMA sequence sets have similar family sizes and correlation bounds being optimal, asymptotically optimal or near-optimal in the Welch Bound sense. Apart from the Kerdock family (and the large Kasami set, which is not discussed) the family size and the correlation bound are all approximately $\sqrt{L}$. L is the sequence length. Another notable feature is the availability of different lengths. Conventional CDMA families have lengths approximating powers of 2 or powers of 4 or 16. Notable exceptions are the Moreno-Tirkel Families A and B whose lengths approximate squares of primes. These features are summarized in Table 2.

| Family | Length - L | Max Correlation | Set Size |
|---|---|---|---|
| **Bent [4]** | $2^{4n}-1$ | $\sqrt{L}$ | $\sqrt{L}$ |
| **Small Kasami [5]** | $2^{2n}-1$ | $\sqrt{L}$ | $\sqrt{L}$ |
| **No-Kumar [6]** | $2^{2n}-1$ | $\sqrt{L}$ | $\sqrt{L}$ |
| **Generalized N-K** | $2^{2n}-1$ | $\sqrt{L}$ | $\sqrt{L}$ |
| **Gold [7]** | $2^n-1$ | $2^{(n+1)/2}-1$ n odd, $2^{(n+2)/2}-1$ n even $\neq 0 \bmod 4$ | $\sqrt{L}$ |
| **Kerdock [8]** | $2\times(2^n-1)$ n odd | $\sqrt{L}$ | L/2 |
| **Moreno-Tirkel A** | p×(p-1) | ≈p | p |
| **Moreno-Tirkel B** | p×(p+1) | ≈p | p |
| **Moreno-Tirkel C** | $2^{2n}-1$ | ≈$2^n$ | $2^n-1$ |

Table 2.

## Conclusions

We compare the new Moreno-Tirkel CDMA sequences with known sequence sets. Our new sequences are the first with non-zero asymptotic normalized linear complexity. We present a conjecture regarding the recursion polynomial of the long sequence. Our construction also delivers new frequency hop patterns. The converse of our construction is introduced. It transforms Kasami/No Kumar sequences into other new frequency hop patterns. However, substituting higher complexity columns in the Kasami/No Kumar arrays does not increase long sequence complexity as in our constructions. This is being investigated.